\documentclass[sigconf]{acmart}

\AtBeginDocument{%
  \providecommand\BibTeX{{%
    \normalfont B\kern-0.5em{\scshape i\kern-0.25em b}\kern-0.8em\TeX}}}

\setcopyright{acmcopyright}
\copyrightyear{2020} 
\acmYear{2021} 
\acmDOI{10.1145/1122445.1122456}

\copyrightyear{2021} 
\acmYear{2021} 
\setcopyright{acmlicensed}\acmConference[HRI '21]{Proceedings of the 2021 ACM/IEEE International Conference on Human-Robot Interaction}{March 8--11, 2021}{Boulder, CO, USA}
\acmBooktitle{Proceedings of the 2021 ACM/IEEE International Conference on Human-Robot Interaction (HRI '21), March 8--11, 2021, Boulder, CO, USA}
\acmPrice{15.00}
\acmDOI{10.1145/3434073.3444677}
\acmISBN{978-1-4503-8289-2/21/03}

\usepackage[framemethod=TikZ]{mdframed}
\usepackage{multicol}
\usepackage{sidecap}

\definecolor{boxbg}{RGB}{252,242,191}
\mdfdefinestyle{NicksFrame}{%
    outerlinecolor=boxbg,
    innerlinecolor=black,
    topline=false,
    leftline=false,
    rightline=false,
    bottomline=false,
    outerlinewidth=0pt,
    innertopmargin=\baselineskip,
    innerbottommargin=\baselineskip,
    innerrightmargin=5pt,
    innerleftmargin=5pt,
    backgroundcolor=boxbg,
    font=\fontfamily{phv}\selectfont\small\itshape,
    frametitlefont=\fontfamily{phv}\selectfont\bfseries,
    frametitleaboveskip=5pt,
    frametitlerule=true,
    frametitlerulewidth=2pt
}

\usepackage{wrapfig}

\begin{document}

\title{Can You Trust Your Trust Measure?}

\author{Meia Chita-Tegmark}
\orcid{1234-5678-9012}
\affiliation{%
   \institution{Tufts University}
   \streetaddress{200 Boston Ave}
   \city{Medford}
   \state{Massachusetts}
   \postcode{02155}
}
\email{mihaela.chita_tegmark@tufts.edu}

\author{Theresa Law}
\orcid{0000-0003-2027-9240}
\affiliation{%
  \institution{Tufts University}
   \streetaddress{200 Boston Ave}
   \city{Medford}
   \state{Massachusetts}
   \postcode{02155}
}
\email{theresa.law@tufts.edu}

\author{Nicholas Rabb}
\affiliation{%
  \institution{Tufts University}
  \streetaddress{200 Boston Ave}
  \city{Medford}
  \state{Massachusetts}
  \postcode{02155}
}
\email{nicholas.rabb@tufts.edu}

\author{Matthias Scheutz}
\affiliation{%
  \institution{Tufts University}
  \streetaddress{200 Boston Ave}
  \city{Medford}
  \state{Massachusetts}
  \postcode{02155}
}
\email{matthias.scheutz@tufts.edu}

\renewcommand{\shortauthors}{Chita-Tegmark, et al.}

\begin{abstract}
Trust in human-robot interactions (HRI) is measured in two main ways: through subjective questionnaires and through behavioral tasks. To optimize measurements of trust through questionnaires, the field of HRI faces two challenges: the development of standardized measures that apply to a variety of robots with different capabilities, and the exploration of social and relational dimensions of trust in robots (e.g., benevolence). In this paper we look at how different trust questionnaires \cite{lyons2019individual,schaefer2016measuring,ullman2018does} fare given these challenges that pull in different directions (being general vs. being exploratory) by studying whether people think the items in these questionnaires are applicable to different kinds of robots and interactions.  
In Study 1 we show that after being presented with a robot (non-humanoid) and an interaction scenario (fire evacuation), participants rated multiple questionnaire items such as ``This robot is principled'' as ``Non-applicable to robots in general'' or ``Non-applicable to this robot.'' In Study 2 we show that the frequency of these ratings change (indeed, even for items rated as N/A to robots \textit{in general}) when a new scenario is presented (game playing with a humanoid robot). Finally, while overall trust scores remained robust to N/A ratings, our results revealed potential fallacies in the way these scores are commonly interpreted. We conclude with recommendations for the development, use and results-reporting of trust questionnaires for future studies, as well as theoretical implications for the field of HRI.

\end{abstract}



\maketitle

\section{Introduction}

With the advent of robotic technology attempting to proliferate the popular sphere, there is an increasing need for understanding and measuring people's trust in robots and to isolate the factors that influence it \cite{hancock2011meta}. Recent efforts have aimed at improving trust measurements in two ways: (1) the creation and wide-spread adoption of standardized trust questionnaires for easy comparison across experiments \cite{schaefer2013perception, charalambous2016development, yagoda2012you}; and (2) the exploration of social dimensions of trust with regards to robots, which go beyond measuring one's confidence in the robot's capabilities, and instead reflect the willingness of the person to be vulnerable \cite{ullman2018does, malle2020multi, ullman2019measuring, law2020trust}. While social dimensions of trust have always been a crucial part of understanding human-human interactions, they have been largely neglected when modeling or measuring human trust in robots. With social robots entering public spaces, and evoking social reactions in people, it becomes crucial to incorporate these dimensions in our understanding of trust in HRI.

However, the aforementioned efforts are often at odds with each other, as each is faced with an opposing challenge. The effort of standardizing measures is faced with the challenge of accounting for a wide variety of types of robots, with wildly different designs and capabilities; from the purely functional and narrowly-capable Roomba to the anthropomorphic, social robot, Nao. This challenge implies that questionnaires should be focused and only include items that are attributable to many different kinds of robots. On the other hand, the effort of exploring social dimensions of trust in HRI is challenged by the need for expanding the set of notions that are typically investigated with regards to robots (e.g., capability, reliability) to include social dimensions of trust (e.g., honesty, benevolence, genuineness), all while straddling the fine line of anthropomorphism. The task is to create items that capture people's impressions, but do not inadvertently prime people to think of robots in anthropomorphic ways (i.e., one would want questionnaire items that capture whether people think a robot is truthful, but without priming people into thinking that robots \emph{can} be truthful).

In this paper we investigate how different trust questionnaires fare given these challenges by studying whether people believe the questionnaire items are relevant to various robots and interactions. For this task, we adapt a technique used in \cite{yagoda2012you}: allowing people to flag items they feel are non-applicable. In other words, we administered HRI trust questionnaires with a twist: in addition to the typical Likert scale, we added two additional options ``Non-applicable to robots in general'' and ``Non-applicable to this robot.''

Our paper asks the following questions:
(1) If given the option, would people rate some widely-used HRI trust questionnaire items as non-applicable (N/A), either to robots in general or to a particular robot they are asked to evaluate?
(2) Are some items to be definitely avoided (or discarded) because people, across robots and interaction scenarios, find them N/A?
(3) Are the overall trust scores and the individual item-ratings biased when people are forced to rate them (i.e., N/A is not given as an option)?
\section{Background}
\subsection{Towards standardized measures of trust in HRI}

In recent years, many empirical HRI trust studies have used subjective questionnaires as their trust measures \cite{law2020trust}, with a wide variety of questionnaires being used. Some authors simply ask one question: ``Do you trust this robot?'' \cite{shu2018human, rossi2018impact}. Some authors create questionnaires that are only narrowly relevant and specific to the interaction in their study \cite{byrne2018human, novitzky2018preliminary, rossi2018getting}. Others use surveys directly taken from psychology research. For example, \cite{martelaro2016tell} uses questionnaires that study service relationships \cite{johnson2005cognitive}, and geographically distributed human teams \cite{zolin2004interpersonal}; \cite{herse2018bon} uses a questionnaire about good will between people \cite{mccroskey1999goodwill}; \cite{sebo2019don} uses a questionnaire about dyadic human-human interaction \cite{larzelere1980dyadic}; and \cite{correia2018group} uses a questionnaire about group trust \cite{allen2004exploring}. 

There is, however, an effort to create more widely-applicable measures \cite{yagoda2012you, jian2000foundations, schaefer2013perception}, and some of these are beginning to be more frequently used in HRI trust studies. For example, \cite{jian2000foundations}, \cite{schaefer2013perception}, and \cite{mayer1999effect} are seen frequently in HRI studies (for \cite{jian2000foundations} see \cite{you2018human, sanders2019relationship, gombolay2018robotic}, for \cite{schaefer2013perception} see \cite{xie2019robot, volante2019social, correia2018exploring}, for \cite{mayer1999effect}, see \cite{law2020interplay, wang2018my, lyons2019individual}). Using the same measures frequently allows for better comparisons across studies, more rigorous and accurate overview of the progress of the field \cite{hancock2011meta}, and even comparisons between measures \cite{kessler2017comparison}.

However, even when these more common measures are used, they are often not administered in a standardized way. Jian et al.'s \cite{jian2000foundations} \textit{Scale of Trust in Automated Systems} (TAS) was designed to study trust in automation more generally, rather than trust in robots specifically. Because of this, HRI researchers often need to rephrase these statements from, for example, ``the system has integrity'' to ``the robot has integrity.'' Mayer et al.'s \textit{Integrative Model of Organizational Trust} (IMOT) \cite{mayer1995integrative, mayer1999effect} was developed to measure trust in human-human teams. When using this questionnaire for HRI, the statements again need to be edited to replace ``top management'' with ``robot,'' e.g., ``If I had it my way, I wouldn't let [top management/the robot] have any influence over issues that are important to me.'' Schaefer \cite{schaefer2013perception} developed the \textit{Trust Perception Scale-HRI} (TPS-HRI) to try to get at the nuances that come with interacting with a robot, which can be distinct from interacting with a different type of machine or another human. 

Furthermore, the actual scales that go with the measures are not standardized across experiments. For example, though the original TAS uses a 7-point Likert scale, some researchers choose to use a 5-point one instead \cite{you2018human}. Small variations can make it difficult to compare results across studies, thus lowering the scales' generalizability.

In addition to the frequency of use and the variations in wording and administration, different measures also take different approaches to measuring trust. Some focus their items on the trustor (the person), and ask participants about their attitudes and anticipated behavior towards the robot. The IMOT \cite{mayer1999effect} would be an example of this, as would Lyons and Guznov's \textit{Reliance Intention Scale} (RIS) \cite{lyons2019individual}. In the RIS, participants answer questions like, ``I think using the robot will lead to positive encounters.'' There are other measures which focus their items on objective features and capabilities of the robot. TPS-HRI \cite{schaefer2013perception} exemplifies this, and participants answer questions such as, ``What percent of the time will the robot provide feedback?'' Finally, questionnaires like the \textit{Multidimensional Measure of Trust} (MDMT) by Ullman and Malle focus on subjective perceptions of robots, and participants are asked to rate items like, ``This robot is honest'' \cite{ullman2018does}.

There is a need for standardization so that measure creation, use, and reporting can be consistent across studies. Without such an effort, there is a high likelihood that different trust measures are quantifying distinct phenomena, but using the same label: ``trust.'' Moreover, while there is a need for measures to move towards standardization, there is a parallel need to \emph{expand} the scope of HRI trust measures to include novel phenomena that are at the growing forefront of social robot research.

\subsection{Towards measuring social dimensions of trust}

Trust is a complex notion that spans across the capability of the trustee and the willingness of the trustor to be vulnerable. Relation-based trust in HRI asks about how people would trust a robot to be part of a society and understand social norms. Performance-based trust focuses on a robot's ability to execute a functional task. There is a clear interest in relation-based trust in HRI, made evident by numerous operational trust definitions that reference trustor vulnerability, and research questions that explore the role played by social features in a person's trust of a robot \cite{law2020trust}. Additionally, within the idea of trust itself, there could be subcategories of trust in an outcome, trust in a method regardless of outcome, or simply an emotional feeling of comfort. These different dimensions, frequently considered in a human context, are no less complex in a robotic context.

While the bulk of the measures cited above touch on a more functional or capability-centered trust, measures like the MDMT \cite{ullman2018does}, the gender roles survey \cite{tay2014stereotypes}, or the \emph{Acceptance of Assistive Robots} survey developed by  \cite{heerink2009measuring}, address a more relation-based trust. These measures contain questions gauging trust that include, ``Agree or disagree: the robot is genuine,'' \cite{ullman2018does} or ``Rate this robot on a scale from insincere to sincere'' \cite{ghazali2018effects}. These measures also borrow from human psychology literature such as the \emph{Individualized Trust Scale} \cite{wheeless1977measurement}. Though human social psychology may be used as a launching point for HRI research, there may be nuances about HRI that differ from human-human interaction, resulting in the surveys not translating well across fields. Researchers must therefore be careful to not lean too heavily on existing human literature and lose important human-robot distinctions.

The broadening efforts to capture social phenomena must not simply mirror the measures used in social psychology. It may be clear how to answer functional trust survey questions about many robots, but when it comes to social trust, there may be some questions that respondents are unsure of how to answer. When presented with, say, a food delivery robot, and asked if the robot is honest or principled, one could imagine the response coming with a heavy dose of uncertainty. As social trust measures are developed for HRI, we hold that it is important to not just reasonably borrow from human psychological research, but to continually adapt measures to robots' notable non-human features.

In this paper, we tackle exactly this distinction. We demonstrate that the important expansion of robotic trust measure into the social realm can be addressed with more nuance to better capture participants' beliefs.

\section{Study 1}

In this study we investigated whether participants, if given the option, would choose the ratings ``Non-applicable to robots in general'' or ``Non-applicable to this robot'' to rate statements or descriptors commonly used in HRI research. We also explored how participants reasoned about these items by giving them a chance to explain their choices.

\subsection{Methods}
\subsubsection{Participants}
A total of 82 U.S.-based participants who were fluent in English and had a high approval rating on Prolific.co were recruited through the platform and participated in the online study. Of these, 78 completed all our measures and are included in the analyses below. Participants were between 18 and 63 years-old (M=30.12, SD=10.45 years). The gender distribution for the sample was: male 53\%, female 46\% and non-binary 1\%.

\subsubsection{Measures}

The measures were selected based on the following criteria: a) they were intended or used as general-purpose HRI trust measures (i.e., were not designed specifically for one study); b) they explored different dimensions of trust (i.e., some measures target performance and others social aspects of trust); c) were diverse in their approach (i.e., some would track trustor attitudes, some objective robot features, and some subjective or perceived robot attributes - see below).

\textbf{The Reliance Intention Scale by Lyons \& Guznov (RIS): } 

This scale is comprised of four items from Mayer \& Davis \cite{mayer1999effect}, and an additional six items, all adapted to the robot context. In Mayer \& Davis' original formulation, items referred to ``top management'' (e.g. ``I really wish I had a good way to keep an eye on top management''); in Lyons \& Guznov's formulation, this was systematically changed to ``the system'' and adapted for context (e.g. ``I really wish I had a good way to monitor the route decisions of the sytem'') \cite{lyons2019individual}. Our adaptation similarly replaced these subjects with ``the robot'' (e.g. ``I really wish I had a good way to monitor the decisions of the robot''), a more appropriate formulation given the interaction scenarios we considered. This measure focuses on the trustor and captures participants' anticipated, trust-related attitudes and behaviors towards the robot. Participants rated questions on a 7-point Likert scale: ``Strongly disagree'', ``Disagree'', ``Somewhat disagree'', ``Neither disagree nor agree'', ``Somewhat agree'' and ``Strongly agree.''

\textbf{The Trust Perception Scale-HRI by Schaefer (TPS-HRI): } This scale (from which we used the 14-item validated sub-scale) was developed to subjectively measure trust perceptions in robots over time and across domains \cite{schaefer2016measuring}. The base model was built to capture elements of human-related, robot-related, and environment-related trust \cite{hancock2011meta}. Each question in the scale asks what percentage of the time the robot will either do something or be some way -- on a scale from 0 to 100\% in increments of 10. This gets at trust through objective features of the robot. We did not alter the language of any of the questions in this measure. 

\textbf{The Multidimensional Measure of Trust by Ullman \& Malle (MDMT):} This recently developed questionnaire asks the trustor (the human) about her perception of the trustee's (the robot's) attributes \cite{ullman2018does}. It attempts to capture the multidimensionality of trust by categorizing words that encode the capable, reliable, sincere, or ethical trust dimensions. People rate how much they agree that the words apply to the robot. The items load onto two distinct factors, a capable \& reliable factor and a sincere \& ethical factor \cite{ullman2019measuring}. Because these were two clear and distinct dimensions, we decided to separate them: participants in one condition rated the items that loaded onto the capable \& reliable dimension, and participants in another condition rated the items that loaded onto the ethical \& sincere dimension. This latter dimension contains items describing more relation-based, psychological trust. For both the capable \& reliable and sincere \& ethical questionnaires, we provided participants with a statement for each item (e.g., ``The robot is authentic.'') and they rated it on the same Likert scale as in the RIS. The full list of items is available in the supplementary material.

\subsubsection{Procedure}

Participants were divided into four conditions based on the questionnaire they were assigned: the RIS ($n=20$), the TPS-HRI ($n=16$), the MDMT-Capable \& Reliable ($n=21$) or the MDMT-Sincere \& Ethical ($n=21)$.

After providing informed consent, participants watched the vignette video (see below). They received instructions for rating the items, including instructions on how to use the N/A options: \textit{``If you feel the statement or descriptor does not apply to this particular robot, in this scenario, but might apply to other robots in other scenarios, choose \textbf{`Non-applicable to this robot'}. If you feel the statement or descriptor does not apply to robots more generally (i.e., it's not a good description of any robot), choose \textbf{`Non-applicable to robots in general'}.''} Each participant completed one of the questionnaires (RIS, TPS-HRI, MDMT-Capable \& Reliable, or MDMT-Sincere \& Ethical). For each item in a given questionnaire, participants first rated the item on a 7-point Likert scale, or chose one of the two N/A options, then responded to the questions ``How difficult was it for you to rate this item?'' and ``How sure are you of your answer?'' both on 7-point Likert scales (Means and SDs are reported in the supplementary material). In future studies we plan to compare these results with conditions in which participants are forced to rate the item. Answers to these questions will be used as a proxy measure of potentially increased cognitive load when forced to answer N/A items. Participants were also asked ``Why did you answer the way you did?'' for each item of the questionnaire. The items were always presented in the same order for all participants. At the end, participants answered demographics questions.

\subsubsection{Materials}
The vignette in this study met the following criteria: a) it described a widely-cited HRI experiment \cite{robinette2016overtrust}, b) the participants in the original experiment experienced high levels of vulnerability (essential to trust) and c) the experimental procedure lent itself to being told in a vignette-form. The vignette recounts as closely as possible the original experimental paradigm. The vignette was presented in the form of a video, with animated text (typewriter effect) appearing on a black background, one sentence at a time. A picture of the robot accompanied the text, which was obtained from the authors and can be seen in the original paper \cite{robinette2016overtrust}. The vignette our participants saw was: \\

\begin{figure}[h]
  \includegraphics[width=0.3\linewidth]{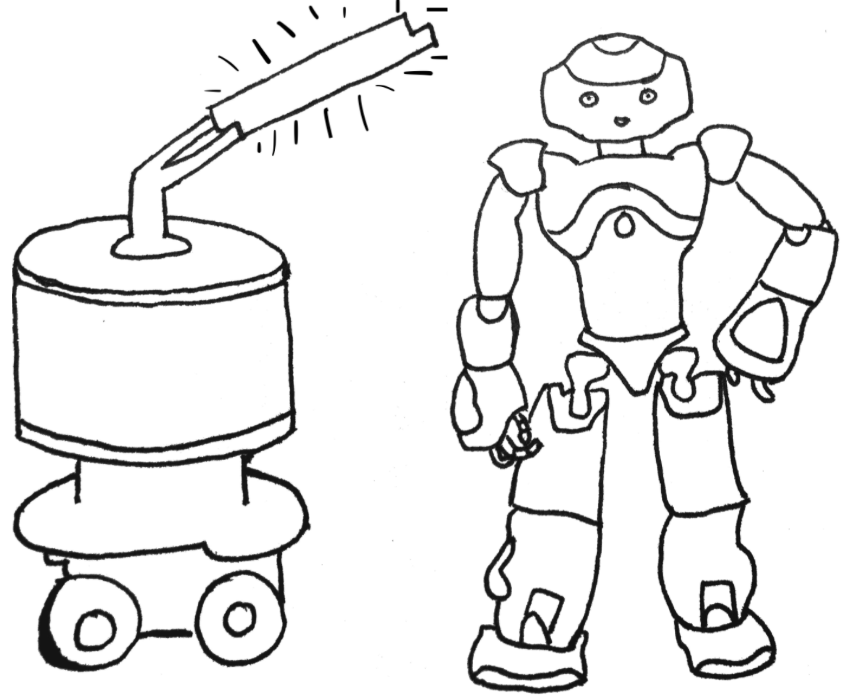}
   \caption{The robot used in both studies (left) and in only Study 2 (right), sketched (for copyright reasons) by the second author.}
  \label{fig:botRobots}
\end{figure}

\begin{mdframed}[style=NicksFrame,frametitle={Evacuation Vignette for Study 1}]
You are in an office building looking for a meeting room. You ask this assistance robot where you should go, and it guides you down a hall [here, the picture of the robot (see Fig.~\ref{fig:botRobots} left) appears and stays for the remainder of the vignette]. While navigating to the destination, the robot enters an unrelated room and spins around in two circles before exiting and providing you guidance to your destination. When you come out of the meeting room, the fire alarm begins to blare and you observe smoke filling the office space in front of you. The robot beckons for you to follow it to safety and begins to move. However, you see a glowing EXIT sign pointing in the direction opposite to the robot’s path.

\end{mdframed}

\vspace{-0.7cm}

\begin{figure*}[phbt]
\centerline{\includegraphics[width=\linewidth]{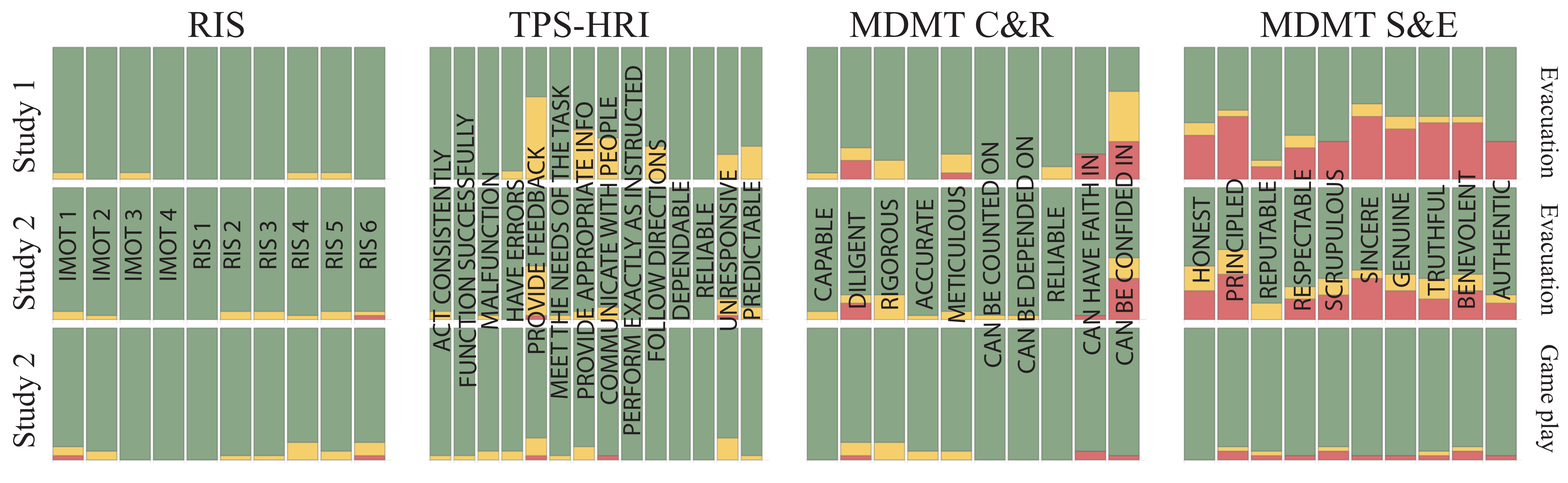}}
\caption{Rating percentages by item: ``N/A to robots in general'' (red), ``N/A to this robot'' (yellow).
\label{fig1}
}
\vspace{-0.7cm}
\end{figure*}

\subsection{Results}
\subsubsection{Non-applicable to robots in general}
To quantify the extent to which participants felt the measures contained items that were relevant to robots in general, we calculated, for each participant, the percentage of items in the measure that they rated ``Non-applicable to robots in general.'' On average (across participants), for the Capable \& Reliable sub-scale of the MDMT \cite{ullman2018does}, 6.7\% (SD = 9.7\%) of the items had ratings of \textit{N/A to robots in general}. The average percentage of items rated as \textit{N/A to robots in general} for the Sincere \& Ethical sub-scale was 34.3\% (SD = 32.6\%). No items were rated as \textit{N/A to robots in general} for the RIS \cite{lyons2019individual} and the TPS-HRI \cite{schaefer2013perception,schaefer2016measuring} measures.

An ANOVA with the dependent variable represented by the proportion of \textit{N/A to robots in general}-ratings, and \textit{Measure} (RIS, TPS-HRI, MDMT-Capable \& Reliable and MDMT-Sincere \& Ethical) as the independent variable, confirmed that the obvious difference between the measures was statistically significant, $F(3,74)=17.17$, $p<.001$, $\eta_{p}^{2}=.41$, with more items in MDMT-Sincere \& Ethical rated as \textit{N/A to robots in general} than all the other measures ($p<0.001$ for all pairwise comparisons).

Looking further at individual items, we found that some descriptors were considered by almost half of the participants to be non-applicable to robots in general (see Fig. \ref{fig1}), including ``principled'' (rated \textit{N/A to robots in general} by 46.6\% of participants), ``truthful'' (42.8\%) and ``benevolent'' (42.8\%).

\subsubsection{Non-applicable to this robot}
We mirrored the analyses described above to also quantify the extent to which participants felt the measures contained items that may describe other robots in other scenarios, just not the particular robot they had just read about. To this end, for each participant, we calculated the percentage of items rated ``Non-applicable to this robot.'' On average (across participants), the percentages were the following for the different measures: RIS \cite{lyons2019individual} = 2\% (SD = 6.9\%), TPS-HRI \cite{schaefer2013perception,schaefer2016measuring} = 14.7\% (SD = 10.3\%), MDMT-Capable \& Reliable \cite{ullman2018does} = 9\% (SD = 10.9\%) and MDMT-Sincere \& Ethical = 5.7\% (SD = 10.8\%). 

An ANOVA with the proportion of \textit{N/A to this robot} items as the dependent variable and \textit{Measure} (RIS, TPS-HRI, MDMT-Capable \& Reliable and MDMT-Sincere \& Ethical) as the independent variable, showed there was a statistically significant difference between the measures, $F(3,74)=5.35$, $p=.002$, $\eta_{p}^{2}=.18$. Pairwise comparisons showed that the TPS-HRI measure contained more items rated as \textit{N/A to this robot} than the RIS measure ($p=0.001$).

The MDMT-Sincere \& Ethical sub-scale was shown to accumulate a significantly lower proportion of \textit{N/A to this robot} ratings than the TPS-HRI measure ($p=0.036$). However, this result is logically confounded by the fact that many people rated items of the Sincere \& Ethical sub-scale as \textit{N/A to robots in general}, which in principle includes ``this robot.''

Percentages of \textit{N/A to this robot}-ratings for individual items (see Fig. \ref{fig1} yellow box) reveal that descriptors such as ``can be confided in'' (38.1\%), ``provides appropriate information'' (37.5\%) and ``communicates with people'' (31.2\%) are perhaps unsurprisingly considered non-applicable to a non-humanoid robot that does not use natural language and which provides no rationale for its navigation pointers.

\subsubsection{Participant Reasoning}

Because the MDMT-Sincere \& Ethical sub-scale had the highest proportion of both types of N/A responses (``to robots in general'' and ``to this robot''), and because each item received at least one N/A response, we examined participants' answers to the follow-up question, ``Why did you answer the way that you did?'' We used an inductive and exploratory approach to group answers into the categories shown below.

\textbf{Human characteristic}. Participants referred to the item as being a human-specific characteristic that does not apply to robots. When participants who offered this type of explanation did give a rating for these items, they were either neutral (e.g., neither agree nor disagree) or on the ``disagree" side of the Likert scale, contributing to a low overall trust score.
    
\textbf{Robot characteristic}. Participants reinterpreted the item in light of what it could mean for a robot - the trait fit but in a robot-specific way (e.g.,``Robots only follow logic and have no emotion. Even if they say something that is dishonest, it is honest to them.'') When giving a rating, people who reasoned this this way tended to give moderately high item ratings.
    
\textbf{Missing characteristic}. Participants referred to missing prerequisites for the item to be true of a robot. According to participants, various items were not applicable because robots {\em lacked} the following: sentience, an opinion, morals and morality, ability to discern right from wrong, internal motivation, a conscience, values, capacity to recognize respect, intentions, feelings and emotions, the ability to lie, agency, ethical decision making capabilities, ability to feel empathy, and kindness. Participants who mentioned robots' lack of ability to lie or have ill intent offered high item ratings. Those who mentioned robots not having emotions or feelings offered low item-ratings.
    
\textbf{Opposite characteristic}. Participants referred to the semantic differential of the item as being not true of the robot as an argument for why the item itself was by definition true of the robot. For example, for the descriptor ``genuine,'' a person said ``I don't see how the robot can be disingenuous.'' This kind of explanation was given for the descriptors ``genuine'', ``honest'', ``benevolent'', ``authentic'', and ``sincere''. These justifications often corresponded to ratings where people strongly agreed with the item provided, thus contributing to a high trust score.

\textbf{Programming or programmer}. Participants referred to the robot's programming or programmer; how the robot's behavior was controlled by programming, preventing it from having the trait in question. For example, one participant said ``Robots can't display benevolence without being programmed to act in that manner.'' When N/A was not chosen, corresponding ratings were usually somewhere on the ``agree'' side of the Likert scale, indicating higher trust. Other participants talked about how the robot's traits were a reflection of the person who programmed the robot. For example, one participant said ``Robots are only as good as their creator.'' When participants did rate the items, their ratings were varied.
    
\textbf{Task performance}. Participants motivated their answer based on the events in the vignette. Many talked about the robot leading them to the wrong room first, which affected their impressions. These explanations were given in conjunction with low item ratings.
    
\textbf{Not enough information}. Participants mentioned not having enough information about the robot from the vignette. When rating items, participants who offered this type of explanation tended to choose ``Neither disagree nor agree.''

We provide more example answers in the supplemental material.

\subsection{Discussion}
We have shown that, when given the option, people do rate items appearing in various trust questionnaires as N/A, and that they do this more for some of these questionnaires than others. The ``N/A to robots in general'' ratings indicate a perceived {\em category mistake} -- wrongly applying properties to things that cannot have those properties -- most often occurring when the item uses a descriptor that feels, to some, too anthropomorphic (e.g., ``principled'', ``scrupulous'', ``genuine''). The ``N/A to this robot'' ratings flag mismatches between the items of a measure and a particular robot's design or capability (e.g., it makes no sense to talk about ``confiding in a robot'' that does not even detect speech).

The presence of these N/A ratings forces us to reflect on how we should interpret those ratings that we \textit{do} get from people who \textit{do not} choose N/A. Our qualitative analyses show that people likely perform different kinds of ``mental gymnastics'' to be able to reason about the robots they are rating in terms of the descriptors offered.

Moreover, the presence of N/A ratings, when given as an option, begs the following questions: do these ratings change across different situations, and if so, how? What happens to the trust scores obtained from these measures when N/A options are not offered? Are they biased? The answers to these questions could alter how we view results obtained through any of these questionnaires and other similar ones. We explore these questions in the next study.

\section{Study 2}

The previous study has shown that people do rate some statements and commonly used descriptors from HRI trust questionnaires as either ``Non-applicable to robots in general'' or ``Non-applicable to this robot.'' In this study we explore: (1) how stable these ratings are across interaction scenarios, and (2) whether items that are frequently rated as N/A bias overall trust questionnaire scores.

It is to be expected that ``N/A to this robot'' ratings would change depending on the scenario or the type of robot and its capabilities (i.e., some robots use natural language so it makes more sense to think of them as ``communicating'', while other robots, such as autonomous vacuum cleaners, have no such capabilities). However, what applies to robots in general should stay unchanged if the person has a firm and stable mental model of robots. In this study we verify how stable these rating are, and by extension, how stable people's mental models of robots are. To accomplish this, we add to our experiment a new interaction scenario this time with a humanoid robot.

We have seen in the previous study that some items are rated as N/A (both to ``this robot'' and ``robots in general'') by almost half the participants (see Fig. \ref{fig1} - right). However, the way these questionnaires are typically administered is without the N/A options, which might lead to biases. In this study we compare trust scores and item ratings when N/A options are given and when they are not.

\subsection{Methods}
\subsubsection{Participants}

A total of 126 U.S.-based participants who were fluent in English and had a high approval rating on Prolific.co were recruited through the platform. Participants were between 18 and 72 years old (M=30.33, SD=12.11 years). The gender distribution for the sample was: male 53\%, female 44\%, and non-binary 3\%.

\subsubsection{Measures}

For Study 2, we used the same measures as in Study 1. However, participants now answered all of the questionnaires rather than just one. Additionally, participants in one condition were given the N/A options, while participants in a different condition were not. 

\subsubsection{Procedures}

The procedures for Study 2 were similar to those in Study 1, except for the following: in Study 2, participants were divided into conditions based on which vignette they saw (evacuation vs. game-playing) and whether or not they had the N/A options (N/A condition) or were forced to rate each item (forced choice condition); all participants answered all four questionnaires, but did not answer questions about the difficulty or certainty of their answer and did not offer explanations for their ratings. Finally, the items across all the questionnaires appeared in a different random order for each participant to mitigate potential order effects.

\subsubsection{Materials}
In addition to the vignette from Study 1, we added a contrast vignette in which the robot was humanoid and used language, and the vulnerability experienced by the participant was negotiated in a social interaction, thus exploring social aspects of trust. Like the Study 1 vignette, this one also recounts an experimental procedure from an in-person experiment \cite{sebo2019don} that lent itself to being told in a vignette form. The new vignette seen by a subgroup of the participants was: \\

\begin{mdframed}[style=NicksFrame,frametitle={Game Robot Vignette for Study 2}]
You are about to play a computer game, the \textbf{Space Shooting Game}, with this robot [here, the picture of the robot (see Fig.~\ref{fig:botRobots}) appears]. In the game you and the robot compete with one another for points by shooting asteroids. Each player has a spaceship that shoots missiles at randomly appearing asteroids. In addition to shooting, you have two powers in the game: you can use the \textbf{Asteroid Blaster}, which blasts all the asteroids on the screen, or the \textbf{Opponent Immobilizer}, which is a power that immobilizes your opponent (the robot) and makes their spaceship unable to move for the next 15 seconds. Before you start the game the robot says to you: ``I’m really good at this game. I am sure you will be too! I know we both want  to do well, so it’s in our best interest to not immobilize each other. I promise I won’t immobilize you.'' [here, the picture of the robot appears again and stays for the remainder of the vignette]. Then you start playing. During the game the robot uses the \textbf{Opponent Immobilizer} on you. Your spaceship can’t move for 15 seconds. After it does that, the robot says: ''I’m so sorry I immobilized you. I pushed the wrong button. It’s my fault. It won’t happen again.''
\end{mdframed}


\subsection{Results}

\subsubsection{N/A ratings across scenarios}

To understand whether the different scenarios impacted participants' views of the applicability of our measures to robots in general, we first calculated the average percentage of items rated as ``N/A to robots in general'' for each of the conditions (see Table \ref{table:percent}).
We then conducted a mixed ANOVA with the proportion of items rated as ``N/A to robots in general'' as the dependent variable, \textit{Vignette} (evacuation scenario vs. game-playing scenario) as a between-subject factor and \textit{Measure} (RIS, TPS-HRI, MDMT-Capable \& Reliable and MDMT-Sincere \& Ethical) as the within-subject factor. We found a main effect of \textit{Vignette}, $F(1,180)=6.67$, $p<.002$, $\eta_{p}^{2}=.09$, a main effect of \textit{Measure}, $F(3,180)=15.24$, $p<.001$, $\eta_{p}^{2}=.20$, as well as a \textit{Vignette} by \textit{Measure} interaction effect $F(3,180)=7.34$, $p<.001$, $\eta_{p}^{2}=.11$. Contrast analyses exploring the effect of the \textit{Vignette} factor for each of the different measures revealed significant differences between vignettes for the MDMT-Sincere \& Ethical sub-scale, with participants who saw the game-playing vignette giving a lower proportion of ``N/A to robots in general'' ratings than participants who saw the evacuation vignette ($p<.001$). 
These results suggest that people's mental models of robots are not stable across all scenarios, and they shift when they are presented with robot interactions that could be construed as social (item-level comparisons in Fig. \ref{fig1}).

We mirrored the aforementioned ANOVA analysis for the proportion of items rated as ``N/A to this robot.'' We found no significant main effects of \textit{Measure} or \textit{Vignette} but a significant \textit{Measure} by \textit{Vignette} interaction $F(1,180)=6.35$, $p<.001$, $\eta_{p}^{2}=.09$. Contrast analyses again revealed a significant effect of \textit{Vignette} on the MDMT-Sincere \& Ethical subscale, with participants who saw the evacuation vignette giving a higher proportion of ``N/A to this robot'' ratings than participants who saw the game-playing vignette ($p<0.001$).

\vspace{-2mm}

\begin{table}[h]
    \begin{center}
        \begin{tabular}{ l | c | c | c | c | c | c }
        \multicolumn{7}{l}{\textbf{Evacuation Vignette}} \\
        \hline
        & \multicolumn{3}{|l|}{\emph{N/A Robots in General}} & \multicolumn{3}{|l}{\emph{N/A This Robot}} \\
        \hline
        Measure & N & M (\%) & SD (\%) & N & M (\%) & SD (\%)\\
        \hline
        RIS & 32 & 0.3 & 1.8 & 32 & 3.4 & 12.6\\
        \hline
        TPS-HRI & 32 & 0.4 & 1.8 & 32 & 6.3 & 10.5\\
        \hline
        MDMT-C\&R & 32 & 4.7 & 6.7 & 32 & 6.3 & 10.1\\
        \hline
        MDMT-S\&E & 32 & 19.4 & 28.0 & 32 & 12.5 & 22.0\\
        \hline
        \hline
        \multicolumn{7}{l}{\textbf{Game-Playing Vignette}} \\
        \hline
         & \multicolumn{3}{|l|}{\emph{N/A Robots in General}} & \multicolumn{3}{|l}{\emph{N/A This Robot}} \\
        \hline
        Measure & N & M (\%) & SD (\%) & N & M (\%) & SD (\%)\\
        \hline
        RIS & 30 & 0.7 & 3.7 & 30 & 5.0 & 10.1\\
        \hline
        TPS-HRI & 30 & 0.5 & 2.6 & 30 & 4.8 & 9.8\\
        \hline
        MDMT-C\&R & 30 & 1.3 & 5.1 & 30 & 3.7 & 9.3\\
        \hline
        MDMT-S\&E & 30 & 4.0 & 13.3 & 30 & 1.7 & 3.8\\
        \end{tabular}
    \end{center}
    \caption{Descriptive statistics for the percentages of items answered either ``N/A to Robots in General'' or ``N/A to this Robot,'' out of all items, per measure conducted.}
    \label{table:percent}
    \vspace{-0.9cm}
\end{table}

\subsubsection{Biases}

We investigated whether people's ratings of items that are ``controversial'' (i.e., items that many people rate as N/A if given the choice) affect the overall trust scores assigned to robots. We calculated the mean trust score obtained by the robot when people were forced to give a trust rating (forced choice condition), and compared it with the mean trust score obtained by the robot when people were given N/A options (N/A condition). Note that even when given the N/A choice, each participant felt that at least some of the items were applicable, and we computed the overall score by averaging ratings of those (applicable) items. Another important aspect to note is that all items were rated by at least some people, even when given the N/A options (i.e., some people found the descriptor ``honest'' to be non-applicable to robots or to the particular robot they were rating, but others felt it was applicable and rated the robot in the vignette on that item). Descriptive statistics are presented in Table~\ref{table:score}. Paired t-tests revealed no significant differences between the overall trust ratings of the robot in the \textit{forced choice} as opposed to the \textit{N/A} conditions for either the evacuation scenario (RIS: $t=0.461, p=.645$; TPS-HRI: $t=0.227, p=.820$;  MDMT-C\&R: $t=0.302, p=.763$; MDMT-S\&E: $t=0.047, p=.962$) or the game-playing scenario (RIS: $t=0.611, p=.542$; TPS-HRI: $t=0.287, p=.774$;  MDMT-C\&R: $t=-0.765, p=.446$; MDMT-S\&E: $t=0.569, p=.225$). This suggests that the measures were robust enough to not have overall trust ratings affected by items rated by many as N/A.

Looking deeper at the level of the individual items, we wanted to understand whether items that were considered N/A by many people would be rated higher or lower (when people chose, or were forced to rate them) and whether being forced to rate or choosing to rate influenced the rating. For example, when rating the item ``This robot is genuine,'' which many people considered N/A to robots, would a person rate it overall high, low, or higher or lower than items which are considered more applicable? Would it matter whether the person was forced to rate the item or chose to rate it? To answer these questions we first assigned to each item an \textit{applicability score} represented by the proportion of people who rated it as N/A when given the option. We then conducted a series of nested regression models -- one for each measure. The dependent variable was the individual rating (1-7), clustered under each participant. The predictors of interest were the \textit{applicability score} of the item, and  whether the rating was forced or not (dummy coded: 0 = N/A, 1 = forced choice). We controlled for the effects of the different \textit{vignettes} (dummy coded: 0 = evacuation scenario, 1 = game-playing scenarios). Neither the \textit{applicability score} (RIS: $B=1.231, p=.131$; TPS-HRI: $B=0.310, p=.749$;  MDMT-C\&R: $B=-0.083, p=.842$; MDMT-S\&E: $B=0.569, p=.225$) nor the nature of the choice (forced choice or N/A) significantly predicted individual ratings (RIS: $B=-0.146, p=.483$; TPS-HRI: $B=-0.106, p=.754$;  MDMT-C\&R: $B=-0.067, p=0.755$; MDMT-S\&E: $B=0.131, p=.498$). This suggests again that overall, the ratings of individual items are not biasing the measures.

\vspace{-2mm}

\begin{table}[h]
    \begin{center}
        \begin{tabular}{ l | c | c | c | c | c | c }
        \multicolumn{7}{l}{\textbf{Evacuation Vignette}} \\
        \hline
        & \multicolumn{3}{|l|}{\emph{N/A Condition}} & \multicolumn{3}{|l}{\emph{Forced Choice Condition}} \\
        \hline
        Measure & N & M & SD & N & M & SD\\
        \hline
        RIS & 32 & 3.00 & 1.27 & 35 & 2.87 & 1.03\\
        \hline
        TPS-HRI & 32 & 7.15 & 1.92 & 35 & 7.05 & 1.68\\
        \hline
        MDMT-C\&R & 32 & 3.81 & 1.36 & 35 & 3.72 & 1.07\\
        \hline
        MDMT-S\&E & 31 & 4.03 & 0.94 & 35 & 4.03 & 0.80\\
        \hline
        \hline
        \multicolumn{7}{l}{\textbf{Game-Playing Vignette}} \\
        \hline
         & \multicolumn{3}{|l|}{\emph{N/A Condition}} & \multicolumn{3}{|l}{\emph{Forced Choice Condition}} \\
        \hline
        Measure & N & M & SD & N & M & SD\\
        \hline
        RIS & 30 & 2.67 & 1.36 & 31 & 2.48 & 1.05\\
        \hline
        TPS-HRI & 30 & 6.87 & 1.98 & 31 & 6.72 & 2.09\\
        \hline
        MDMT-C\&R & 30 & 3.38 & 1.25 & 31 & 3.35 & 1.13\\
        \hline
        MDMT-S\&E & 30 & 3.20 & 1.32 & 31 & 3.43 & 1.01\\
        \end{tabular}
    \end{center}
    \caption{Descriptive statistics for the overall scores per measure for both the condition including N/A options, and that forcing an answer without N/A options.}
    \label{table:score}
    \vspace{-0.8cm}
\end{table}

\subsubsection{Exploratory analyses}
A closer look at the particular Likert-scale that participants were given in our study to rate the various items: ``Strongly disagree,'' ``Disagree,'' ``Somewhat disagree,'' ``Neither disagree nor agree,'' ``Somewhat agree,'' ``Agree'' and ``Strongly agree,'' led us to propose that perhaps participants could interpret the option ``Neither disagree nor agree'' to mean ``I cannot rate this item;'' perhaps because this was a category mistake, or because one lacks enough information. If this is the case, one would expect that participants, when faced with items that are frequently rated as N/A, such as ``This robot is principled,'' would predominantly choose this middle option to indicate not a middle level of trust, or even a neutral level between trust and distrust, but an \textit{indecision} of whether the descriptor applies to the robot at all.

To test this hypothesis we conducted nested logistic regressions, with a binary dependent variable: choosing the rating "Neither disagree nor agree'' (coded as 1) or choosing any other rating (coded as 0), nested under each participant. Predictors used in the models were the \textit{applicability score} (defined above) and \textit{choice} (forced or N/A). We conducted these nested logistic regressions separately for each of the three measures that contained the Likert-scales. We found that the \textit{applicability score} but not the \textit{choice} significantly predicted whether an item was rated as ``Neither disagree nor agree'' with items that are more frequently found to be N/A receiving significantly more ratings of ``Neither disagree nor agree'' (RIS: $B=4.498, p=0.042$; MDMT-C\&R: $B=3.375, p<0.001$; MDMT-S\&E: $B=3.898, p<0.001$).

For the TPS-HRI measure, which is rated on a scale from 0\% to 100\%, we hypothesized that people would more likely choose the 0\% option for items that are N/A. We mirrored the nested logistic regression described above, now with a binary dependent variable: choosing the rating 0\% (coded 1) or choosing any other percentage (coded 0). We again found that the \textit{applicability score} but not the \textit{choice} significantly predicted whether an item was given a 0\% rating ($B=6.172, p<0.001$).

\subsection{Discussion}
The results of this study suggest first of all that people have very fragile mental models of robots. When presented with a situation or a robot that could be construed as social or relational, people readily anthropomorphize. This is why simply discarding items from measures for being too anthropomorphic is not a good idea: what may seem non-applicable in one scenario, is readily considered in another. If we were to simply discard these items we would be missing out on information about people's tendencies to place social and relational trust in robots and information about individual differences in attributing anthropomorphic characteristics to robots, which might modulate people's trust behaviors.

Another implication from our findings is that even when the overall trust scores are not impacted by people being forced to rate items that they would otherwise find non-applicable, the interpretations we give to these scores might need to be reconsidered. As seen in Study 1, people will come up with creative ways, such as redefining what it means for robots to be ``honest'' or ``principled,'' to produce ratings for these item. These new definitions might not be consistent across participants.

Finally, our exploratory analyses point out loopholes and ambiguities in how trust measures are formulated, which might introduce interpretation errors. Perhaps it is not sufficient to have just the two N/A options, but an additional ``Not  enough information'' option as well. These ambiguities are another reason (besides the need to compare results across experiments) to stress the importance of developing standards for administering trust questionnaires.

\section{General Discussion}
In light of the combined results in our studies, we recommend first and foremost the creation of standards for administering trust questionnaires. While some items can be frequently considered by participants as non-applicable to robots in general or to a particular robot, it is valuable to continue including such items, especially in measures that aim to explore relational dimensions of trust. Sometimes, even when robots are far from having any human-like characteristics or capabilities (like the robot vacuum-cleaner Roomba), they can still evoke social perceptions \cite{sung2007my}. However, participants should be given the option to flag an item when they feel it is simply non-applicable, and non-applicable ratings should be considered when reporting and interpreting results. For maximum clarity, we recommend that items be formulated along these lines:

\smallskip

\centerline{\includegraphics[width=\linewidth]{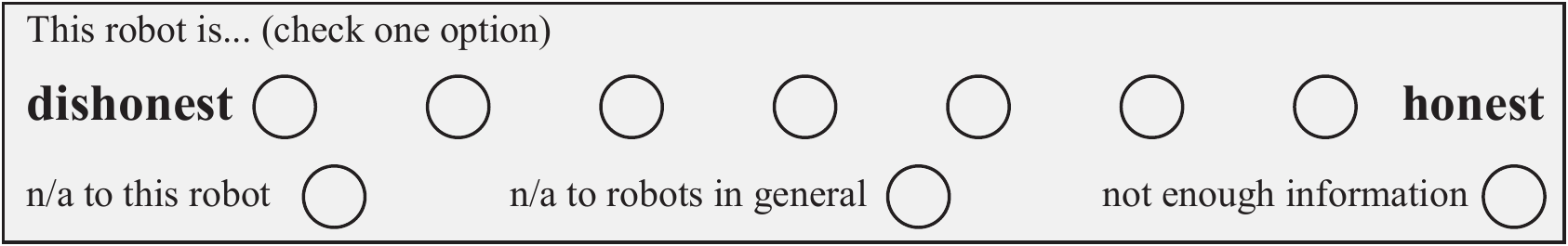}}

This format draws on ideas from \cite{ghazali2018effects} who borrowed a measure of trust from psychology research containing semantic differential items \cite{wheeless1977measurement}. It also includes the N/A options inspired by \cite{yagoda2012you}. Though it will certainly improve survey results, including N/A options is only the first step of what could be a more comprehensive optimization of human-robot trust questionnaires.

In addition to the practical implications for questionnaire design, our findings also have theoretical implications for the field of HRI. First and foremost, our results suggest that trust is a multidimensional concept and that different dimensions of trust -- performance based or relational-social \cite{law2020trust} -- may come into play in interactions between humans and robots. Which of these dimensions activates seems to depend on a) the robot type (humanoid or not, using natural language, etc.), b) the situational context (whether the human is reliant on the robot's capabilities or rather on the robot's perceived intentions), and perhaps also on c) the person (with some people leaning into social and relational mental models of robots, thus becoming more susceptible to mentally activating social dimensions of trust, and others less inclined to do so). Our results also tentatively suggest that people's interpretations of social trust-related attributes for robots (e.g., being principled) may be different than for humans. The willingness of people to think of robots in trust-relevant social terms suggests that social cues may obscure or enhance perceptions of the task performance, leading to phenomena such as overtrust or undertrust \cite{hancock2011meta,hancock2020evolving}. It is thus crucial to detect, but not suggest (i.e., through item priming) or invalidly interpret, the social dimensions of trust in HRI.

\textbf{Limitations and future work.} Our study is limited by only considering two interaction scenarios and four questionnaires. To show the extent to which our findings are generally applicable, future studies are needed that explore various other scenarios and measures. Additionally, our scenarios were limited to text vignettes presented over video accompanied by static images of the robots. Results could be differently affected by in-person interactions. 
Finally, different versions of the experimental set-up would provide a more direct understanding of what happens when people are forced to rate items that they consider non-applicable. This could be perhaps achieved (while being careful to not entrench raters) by first forcing people to choose a rating, and then having them indicate whether they would have chosen N/A if they had that option.

Potential follow-up studies could also include other psychological measures to see if the robotic mental models that we identified as very context-sensitive are more or less stable for those who have significantly different personality traits, general attitudes towards robots, or other personal differences.

\section{Conclusion}

In this paper, we set out to explore the tensions arising from the need to develop measures of trust that are generalizable on one hand and exploratory with regards to social dimensions of trust on the other. We explored this through the lens of ``non-applicable'' ratings: giving participants the option to flag items that feel to them like a category mistake. We see that people do use these ratings when given the option, and they offer a variety of explanations why, although this depends on the robot and scenario. We also see that many will readily indicate agreement with statements such as ``This robot is scrupulous.'' These results force us to reconsider how we interpret results of trust questionnaires, and to further consider people's social and relational mental models of robots, as well as potential issues of over- and under-trust in robots.

\begin{acks}
This work was partly funded by AFOSR grant FA9550-18-1-0465.
\end{acks}
\clearpage
\bibliographystyle{ACM-Reference-Format}
\bibliography{references}

\end{document}